# Twistoptics in Planar Heterostructures with an Arbitrary Number of Rotated 3D Thin Layers and 2D Conductive Sheets

Christian Lanza[1,2,†,*], José Álvarez-Cuervo[1,2,†], Kirill V. Voronin[3], Gonzalo Álvarez-Pérez[4], Aitana Tarazaga Martín-Luengo[1,2], Javier Martín-Sánchez[1,2], Alexey Y. Nikitin[3,5,*], and Pablo Alonso-González[1,2,*]

[1]*Department of Physics, University of Oviedo, Oviedo, Spain.*
[2]*Center of Research on Nanomaterials and Nanotechnology CINN (CSIC–Universidad de Oviedo), El Entrego, Spain.*
[3]*Donostia International Physics Center (DIPC), Donostia/San Sebastián 20018, Spain.*
[4]*Center for Biomolecular Nanotechnologies, Istituto Italiano di Tecnologia, via Barsanti 14, Arnesano, 73010, Italy.*
[5]*IKERBASQUE, Basque Foundation for Science, Bilbao 48013, Spain.*

[†]These authors contributed equally to this work
*lanzachristian@uniovi.es; alexey@dipc.org; pabloalonso@uniovi.es

**Twistoptics has recently emerged as a branch of nano-optics that explores light propagation in stacks of thin anisotropic layers rotated relative to one another. The concept is particularly relevant for polaritons –hybrid light-matter quasiparticles- in van der Waals (vdW) materials, where strong in-plane anisotropy and deep subwavelength confinement make the polaritonic dispersion highly sensitive to interlayer twist angles. This sensitivity enables exotic phenomena such as canalization, i.e., diffraction-free propagation, with potential applications ranging from thermal management to super-resolution imaging. Despite rapid progress, a general analytical framework to describe polariton propagation in twisted planar heterostructures has been missing. Here we present an analytical model for planar stacks comprising an arbitrary number of finite-thickness anisotropic (biaxial) layers and infinitesimally thin anisotropic conductive sheets. The formalism and its high-momentum and thin-film approximations predict key polaritonic observables, such as wavelength, propagation length, and electromagnetic field distributions. We also provide open-access numerical scripts implementing the model to support their practical use. Together, these results provide a general theoretical foundation for twistoptics and should facilitate the discovery and accelerate the implementation of twist-engineered polaritonic phenomena across the electromagnetic spectrum.**

## I. INTRODUCTION

The advent of van der Waals (vdW) materials [1,2], crystals assembled from stacks of atomically thin layers held together by weak out-of-plane vdW forces, has enabled the development of *twistronics* [3], in which electronic properties are engineered by controlling the relative twist angle between stacked layers. This structural degree of freedom generates Moiré superlattices that host a range of emergent quantum states, including unconventional superconductivity and correlated insulator phases [4–7]. Building on these advances, the same concept has recently been expanded to *twistoptics* [8,9], which exploits interlayer rotations to tailor light-matter interactions at the nanoscale [10-14]. A prominent example is the control of in-plane anisotropic polaritons supported by biaxial vdW crystals [15-18], enabling phenomena such as canalization [19,20], i.e., strongly directional, nearly diffraction-free propagation along a single in-plane axis. This effect was first demonstrated in twisted α-MoO$_3$ bilayers [21–24] and trilayers [25].

To date, theoretical studies of such twisted heterostructures have relied primarily on the Transfer Matrix Method (TMM) [26] and full-wave numerical simulations, such as the Finite Elements Method (FEM). These approaches accurately reproduce experimental dispersion relations and near-field images, but the field still lacks a general analytical framework, particularly for hybrid platforms combining an arbitrary number of biaxal crystals with 2D conductive sheets such as graphene [27-29]. Existing analytical treatments are often restricted to two or three layers or rely on thin-film approximations [24], which can become inaccurate depending on the flake thickness, spectral range, or twist angle. Consequently, the absence of a general, material-independent analytical model hinders a priori predictions and hampers the rapid optimization of diverse twisted multilayer heterostructures.

Here, we address these limitations by developing a general analytical framework for electromagnetic quasi-normal modes in heterostructures composed of an arbitrary number of rotated biaxial layers and 2D conductive sheets. The theory provides analytical access to key polaritonic properties, including wavelength, propagation length, and electromagnetic field distributions. In particular, we derive compact expressions for the dispersion relation, the electromagnetic field amplitudes within each constituent layer, and the $p$-polarized out-of-plane Fresnel reflection coefficients. These results enable the reconstruction of polaritonic electric field modes across arbitrary cross-sections of the stack, reproducing numerical calculations (i.e., TMM [26]) or full-wave simulations while offering explicit control over the underlying physical degrees of freedom and significantly improving computational efficiency. Finally, we use the Dyadic Green's Function (DGF) formalism [30, 31] to model polariton propagation excited by a point dipole source without resorting to full-wave simulations. Together with open-access numerical scripts, this work establishes a robust analytical foundation for twistoptics, with engineering capabilities useful for applications spanning imaging, radiative and thermal management [32], and inverse design [33–35].

## II. GENERAL DISPERSION OF POLARITONS IN HETEROSTRUCTURES OF 3D BIAXIAL CRYSTALS AND 2D CONDUCTIVE SHEETS

The optical response of a crystal with permeability tensor $\hat{\boldsymbol{\mu}} = \hat{\boldsymbol{I}}$, where $\hat{\boldsymbol{I}}$ is the identity matrix, is generally described by its complex-valued permittivity tensor $\hat{\boldsymbol{\varepsilon}}$. In the most general case, $\hat{\boldsymbol{\varepsilon}}$ is non-diagonal. For example, in monoclinic and triclinic crystals the real and imaginary parts of $\hat{\boldsymbol{\varepsilon}}$ cannot be simultaneously diagonalized in the same coordinate system [36–41]. By contrast, for orthorhombic crystals (and crystals with higher symmetry), both $\text{Re}(\hat{\boldsymbol{\varepsilon}})$ and $\text{Im}(\hat{\boldsymbol{\varepsilon}})$ can be diagonalized in their principal coordinate system, which we denote by $\{x, y, z\}$. In this frame, the permittivity tensor takes the form

$$\hat{\boldsymbol{\varepsilon}} = \begin{pmatrix} \varepsilon_x & 0 & 0 \\ 0 & \varepsilon_y & 0 \\ 0 & 0 & \varepsilon_z \end{pmatrix}, \tag{1}$$

where for biaxial crystals, in general, $\varepsilon_x \neq \varepsilon_y \neq \varepsilon_z$ [15,16]. Depending on the values of the diagonal elements in Eq. (1), the crystal couples differently to the electromagnetic field along different crystallographic directions. In particular, some anisotropic materials exhibit opposite signs among the diagonal components of $\text{Re}(\hat{\boldsymbol{\varepsilon}})$ (i.e., at least one component has a sign opposite to the others) within specific spectral windows, which enables hyperbolic polaritons. This behavior has been recently encountered in the visible range for plasmon polaritons (PPs) in MoOCl$_2$ [42,43], and in the infrared and terahertz ranges for phonon polaritons (PhPs) in polar dielectrics [10-16], within the so-called Reststrahlen bands [44].

In contrast, the optical response of a 2D material is conveniently described by its surface conductivity tensor $\hat{\boldsymbol{\sigma}}$, which in the principal in-plane axes $\{x, y\}$ can be written as

$$\hat{\boldsymbol{\sigma}} = \begin{pmatrix} \sigma_x & 0 \\ 0 & \sigma_y \end{pmatrix}. \tag{2}$$

In the following, we consider a heterostructure composed of $N$ rotated, finite-thickness biaxial layers ("3D layers") embedded between two isotropic semi-infinite media (superstrate and substrate). We further assume that a 2D conductive sheet is located at each interface between adjacent finite-thickness layers, giving a total of $N + 1$ sheets, as schematically shown in Fig. 1a.

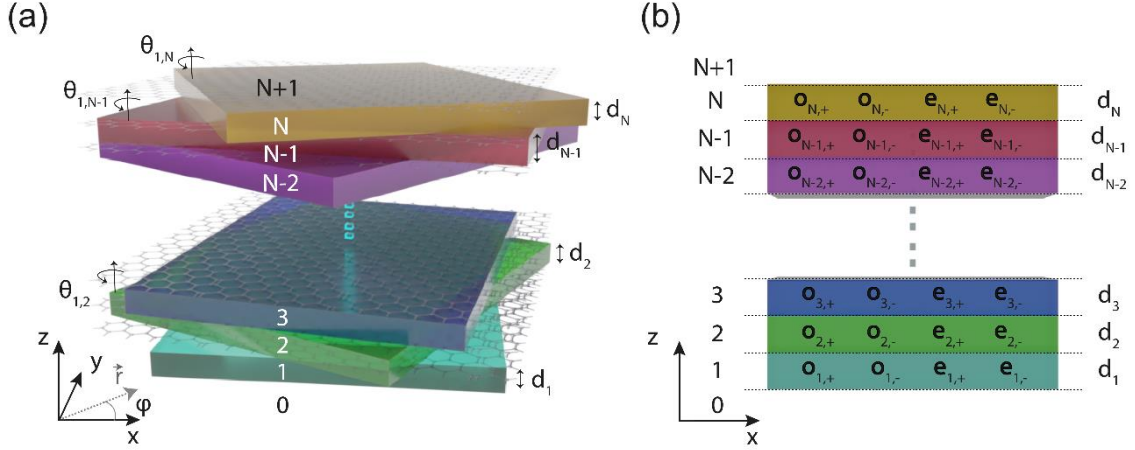

FIG. 1. Schematic of a heterostructure composed of $N$ rotated thin 3D biaxial layers and $N + 1$ anisotropic 2D conductive sheets. (a) 3D sketch of the heterostructure. The coordinate system is aligned to the principal crystal axes of the bottom layer. The angle $\varphi$ denotes the in-plane polariton propagation direction in this reference frame. The thickness of the j-th layer is $d_j$. The twist of each biaxial layer relative to the bottom layers is parameterized by $\theta_{1,j}$. An anisotropic 2D conductive sheet is present at each interface; for simplicity, their twist angles are not shown. The gap between the top and bottom parts of the drawing indicates the omitted intermediate 3D layers and 2D sheets. (b) Cross-sectional view of the heterostructure. The coordinate axes are aligned with the $x$ and $z$ crystal directions of the bottom layer. The electromagnetic fields within each layer are expressed as a linear combination of upgoing and downgoing ordinary and extraordinary waves.

The angle $\varphi$ denotes the in-plane polar angle in our coordinate frame, defined with respect to the crystallographic axes of the bottom layer. Each 3D layer has thickness $d_j$ and is rotated by a twist angle $\theta_{1,j}$ relative to the bottom layer. The 2D conductive sheets are rotated by angles $\phi_{jk}$, also referenced to the bottom layer, where $j$ (with $k = j - 1$) labels the j-th (and k-th) 3D layers; thus, the $jk$ sheet lies at the interface between the j-th and k-th layers. For convenience, we choose the interface between the bottom layer ($j = 1$) and the substrate ($k = 0$) to coincide with the plane $z = 0$.

To describe the electromagnetic fields in the heterostructure, we expand them within each layer as a superposition of plane-waves as

$$\mathbf{E} = E_0 \mathbf{e}\, e^{i(\mathbf{k}\cdot\mathbf{r}-\omega t)}, \qquad \mathbf{H} = H_0 \mathbf{h}\, e^{i(\mathbf{k}\cdot\mathbf{r}-\omega t)}, \tag{3}$$

where $\mathbf{e}$ and $\mathbf{h}$ are the field basis vectors, $E_0$ and $H_0$ are field amplitudes, $\omega$ is the angular frequency, $\mathbf{k}$ is the complex wavevector, and $\mathbf{r}$ is the position vector. The electric-field basis $\mathbf{e}$ follow from Maxwell's equations:

$$\nabla(\nabla \mathbf{E}) - \Delta \mathbf{E} = k_0^2 \hat{\varepsilon} \mathbf{E}, \tag{4}$$

where $k_0 = \frac{\omega_0}{c}$ is the free-space wavenumber. The characteristic equation associated with Eq. (4) yields two non-trivial solutions $q_{\gamma,z}^2 = \frac{k_{\gamma,z}^2}{k_0^2}$ given by [45,46]:

$$q_{\gamma,z}^2 = \frac{1}{2}\left[\frac{\varepsilon_x + \varepsilon_z}{\varepsilon_z}q_x^2 + \frac{\varepsilon_y + \varepsilon_z}{\varepsilon_z}q_y^2 - \varepsilon_x - \varepsilon_y\right] \pm \frac{1}{2}\sqrt{D}, \tag{5}$$

where $\gamma = e_1, e_2$ labels the two extraordinary waves supported inside a biaxial medium [45], and $q_{x,y} = \frac{k_{x,y}}{k_0}$ are the normalized in-plane components of the wavevector. The discriminant reads

$$D = \left(\varepsilon_x - \varepsilon_y + \frac{\varepsilon_z - \varepsilon_x}{\varepsilon_z}q_x^2 - \frac{\varepsilon_z - \varepsilon_y}{\varepsilon_z}q_y^2\right)^2 + 4\frac{(\varepsilon_z - \varepsilon_x)(\varepsilon_z - \varepsilon_y)}{\varepsilon_z^2}q_x^2 q_y^2. \tag{6}$$

Equation (5) highlights that the out-of-plane wavevector depends on both the permittivity tensor components and the in-plane momentum $(q_x, q_y)$. Explicit expressions for the basis vectors $\mathbf{e}_1$ and $\mathbf{e}_2$ are provided in Refs. [47,48].

For a given j-th layer of the heterostructure, the electric field can be written as a superposition of the corresponding extraordinary plane waves $(\mathbf{e}_{1,j}, \mathbf{e}_{2,j})$,

$$\mathbf{E}_j = \sum_{\gamma_j = \mathbf{e}_{1,j}, \mathbf{e}_{2,j}} \left(a_{\gamma_j}^{\uparrow} \boldsymbol{\gamma}_{j,+} e^{i\xi_{j,\gamma,z}} + a_{\gamma_j}^{\downarrow} \boldsymbol{\gamma}_{j,-} e^{-i\xi_{j,\gamma,z}}\right), \tag{7}$$

where $a_{\gamma_j}^{\uparrow}$ and $a_{\gamma_j}^{\downarrow}$ are the complex amplitudes of the upgoing and downgoing waves, respectively (Fig. 1b), and $\xi_{j,\gamma,z} = k_{j,\gamma,z} z$ is the phase with $k_{j,\gamma,z} = iq_{j,\gamma,z} k_0$. In $\boldsymbol{\gamma}_{j,\pm}$, the $+$ ($-$) sign corresponds to propagation along (opposite to) the $z$ axis.

In the semi-infinite isotropic substrate and superstrate (regions 0 and $N + 1$, respectively), the normalized out-of-plane component satisfies $q_z = \sqrt{q^2 - \varepsilon}$, where $\varepsilon$ is the scalar dielectric permittivity and $q^2 = q_x^2 + q_y^2$. In this limit, $\mathbf{e}_1$ and $\mathbf{e}_2$ reduce to the standard basis vectors for the $s$- and $p$-polarized waves, respectively [47]. The electric fields in these regions can be written as

$$\begin{aligned}\mathbf{E}_0 &= \sum_{\beta=s,p} a_{\beta}^0 \boldsymbol{\beta}_{0,-} e^{-i\xi_{0,z}} \\ \mathbf{E}_{N+1} &= \sum_{\beta=s,p} a_{\beta}^{N+1} \boldsymbol{\beta}_{N+1,+} e^{i\xi_{N+1,z}}\end{aligned}, \tag{8}$$

where $a_{\beta}^i$ is the corresponding amplitude and $\xi_{i,z} = q_{i,z} k_0 z$ is the phase in the substrate ($i = 0$) and superstrate ($i = N + 1$). Here $\boldsymbol{\beta}_{0,-}$ ($\boldsymbol{\beta}_{N+1,+}$) denotes $s$- and $p$- polarized waves propagating opposite to (along) the $z$-axis within the substrate (superstrate).

Using Faraday's law, $\nabla \times \mathbf{E} = -\frac{1}{c}\frac{\partial \mathbf{H}}{\partial t}$, together with Ohm's law $\mathbf{J} = \hat{\sigma}\mathbf{E}$, the magnetic field in the j-th layer and the surface current at the $jk$ interface can be written as:

$$\begin{aligned}\mathbf{H}_j &= \mathbf{q}_j \times \mathbf{E}_j \\ \mathbf{j}_{jk} &= \frac{c}{2\pi}\hat{\alpha}_{jk}\mathbf{E}_{j,t}\end{aligned}, \tag{9}$$

where the subscript $jk$ refers to the 2D conductive sheet located between media $j$ and $k = j - 1$, $\mathbf{E}_{j,t}$ is the tangential electric field at the interface, and $\hat{\alpha} = \frac{2\pi}{c}\hat{\sigma}$ is the normalized optical conductivity [49].

To determine the electromagnetic quasi-normal modes supported by the heterostructure (orthogonal modes are not expected in the presence of losses), we enforce the boundary conditions at each interface: (i) continuity of the tangential electric field and (ii) the discontinuity of the tangential magnetic field due to surface currents associated with the 2D sheet. At the interface between media $j$ and $k$, these conditions read

$$\mathbf{E}_{j,t} = \mathbf{E}_{k,t}$$
$$\mathbf{e}_z \times [\mathbf{H}_j - \mathbf{H}_k] = 2\hat{\alpha}_{jk}[\mathbf{e}_z \times \mathbf{E}_k] \tag{3}$$

which form a set of algebraic equations that can be cast in compact matrix form as

$$M\mathbf{a} = \mathbf{0}, \tag{4}$$

where the vector $\mathbf{a} = \left(a^{\uparrow}_{\mathbf{e}_{1,N}} \quad a^{\downarrow}_{\mathbf{e}_{1,N}} \quad a^{\uparrow}_{\mathbf{e}_{2,N}} \quad a^{\downarrow}_{\mathbf{e}_{2,N}} \quad \cdots \quad a^{\uparrow}_{\mathbf{e}_{1,1}} \quad a^{\downarrow}_{\mathbf{e}_{1,1}} \quad a^{\uparrow}_{\mathbf{e}_{2,1}} \quad a^{\downarrow}_{\mathbf{e}_{2,1}}\right)^T$ collects the unknown plane-wave amplitudes in each layer (cf. Eq. (7)). The $4N \times 4N$ matrix $M$ can be compactly expressed as

$$M = \begin{pmatrix} D_{N+1,N} & 0 & 0 & 0 & 0 \\ D^L_{N,N-1} & D^R_{N,N-1} & 0 & 0 & 0 \\ 0 & \cdots & \cdots & 0 & 0 \\ 0 & 0 & D^L_{j,k} & D^R_{j,k} & 0 \\ 0 & 0 & 0 & \cdots & \cdots \\ 0 & 0 & 0 & 0 & D_{1,0} \end{pmatrix}. \tag{5}$$

A detailed derivation of Eq. (4), together with explicit expressions for the constituent elements of $M$, is provided in the Supplementary Information (Ref. [48]).

The electromagnetic quasi-normal modes of the heterostructure are determined by the non-trivial solutions of Eq. (4), which correspond to the zeros of the determinant $|M|$ in Eq. (5). In the general case, $|M|$ does not admit a simple closed-form expression, as the number of terms in the associated polynomial scales as $4N!$. Nevertheless, in certain limiting cases -for instance, a single biaxial layer ($N = 1$) without 2D sheets- analytical solutions can be obtained under two widely used approximations [47]: (i) the high-momentum (or large-refractive-index) approximation, appropriate for strongly confined polaritons in individual biaxial layers, and (ii) the thin-film approximation, where the biaxial layer is sufficiently thin to be modeled as an effective 2D conductive sheet. Both approximations have been shown to accurately describe polaritons in thin vdW layers across broad parameter ranges of practical interest. In the following sections, we adopt these two approximations to derive tractable dispersion relations for the more complex heterostructures considered here (Fig. 1).

### III. HIGH-MOMENTUM APPROXIMATION IN ARBITRARILY LARGE HETEROSTRUCTURES

The high-momentum approximation assumes a large normalized in-plane polariton wavevector; i.e., $|q| \gg 1$. In this limit, the normalized out-of-plane wavevector components in Eq. (5), associated with the two extraordinary waves in each biaxial layer, simplify to [47]

$$q_{j,e_1,z} = q$$
$$q^2_{j,e_2,z} = \frac{\varepsilon_{j,x}}{\varepsilon_{j,z}} q^2_{j,x} + \frac{\varepsilon_{j,y}}{\varepsilon_{j,z}} q^2_{j,y}. \tag{6}$$

Notably, under this approximation $q_{j,e_1,z}$ reduces to the out-of-plane wavevector of an ordinary mode in a uniaxial medium. Henceforth, we refer to the mode associated with $q_{j,e_1,z}$ as the ordinary mode (subscript "$o$".), while the mode associated with $q_{j,e_2,z}$ is retained as the extraordinary mode (subscript "$e$").

Using Eq. (6), we evaluate the zeros of $|M|$ of Eq. (5) by retaining the highest order terms in $q$. The detailed derivations for the case $N = 2$ (a twisted bilayer with $N + 1 = 3$ 2D conductive sheets at the interfaces) and the generalization to arbitrary $N$ are provided in Ref. [48]. In brief, $|M|$ factorizes into the product of the determinants of two $2N \times 2N$ submatrices, one associated with ordinary modes ($M_o$) and the other with extraordinary waves ($M_e$), such that the condition $|M| = 0$ becomes

$$|M| = \begin{vmatrix} M_o & 0 \\ 0 & M_e \end{vmatrix} = |M_o||M_e| = 0. \tag{7}$$

As shown in Ref. [48] (and in Ref. [47] for the particular case $N = 1$ without 2D conductive sheets), the condition $|M_o| = 0$ does not yield a physical solution within the high-momentum regime; consequently, the confined solutions are governed solely by the equation $|M_e| = 0$, meaning that it provides the dispersion of highly confined polaritons. For an arbitrary number $N$ of twisted 3D layers and 2D conductive sheets, this dispersion relation can be written compactly as:

$$\mathcal{F}_N(\mathbf{q}; \omega) \equiv \sum_{l_1=0}^{1} \sum_{l_2=0}^{1} \cdots \sum_{l_N=0}^{1} \left[ \mathcal{C}(l_1, \ldots, l_N) \prod_{i=1}^{N} (\tanh \xi_i)^{l_i} Q_i^{1-l_i} \right] = 0, \tag{15}$$

$$\mathcal{C}(l_1, \ldots, l_N) = \begin{pmatrix} 1 \\ Q_{N+1} + I_{N+1} \end{pmatrix}^T \prod_{j=1}^{N} \left[ \begin{pmatrix} 0 & Q_j^2 \\ 1 & I_j \end{pmatrix} l_j + \begin{pmatrix} 1 & I_j \\ 0 & 1 \end{pmatrix} (1 - l_j) \right] \begin{pmatrix} Q_0 \\ 1 \end{pmatrix}.$$

where $Q_j = \varepsilon_{j,z} q_{j,e,z}$, $\xi_j = q_{j,e,z} k_0 d_j$ and $I_j = 2i\alpha_{jk,eff} q^2$ with $\alpha_{eff} q^2 = \alpha_{xx} q_x^2 + 2\alpha_{xy} q_x q_y + \alpha_{yy} q_y^2$.

Solving Eq. (15) yields the normalized, generally complex, in-plane momentum $\mathbf{q}(\omega)$ of the electromagnetic modes supported by the twisted multilayer heterostructure at a given frequency $\omega$. This momentum determines key modal characteristics (i.e., wavelength and propagation length). To reconstruct the full electromagnetic fields of the corresponding quasi-normal eigenmodes -and their spatial evolution- one must additionally obtain the plane-wave amplitudes entering the field expansions in Eq. (7) and Eq. (8). These amplitudes follow from Eq. (11) as the non-trivial null space of $M$ evaluated at the solution $\mathbf{q}(\omega)$, up to an arbitrary overall normalization.

The functions $Q_j$ and $\xi_j$ contain all the relevant information of the j-th 3D layer, including its permittivity tensor, twist angle, and thickness, whereas $I_j$ accounts for the interfacial $jk$ 2D conductive sheet through its conductivity tensor and twist angle. In addition, two 2D conductive sheets $i$ and $j$, stacked at the same interface, contribute additively and can be treated as a single effective sheet, $I = I_i + I_j$. Finally, the contribution of the j-th 3D layer can be removed by setting $d_j = 0$, while the effect of the $jk$ 2D conductive sheet is suppressed by setting $I_j = 0$.

To illustrate the practical use of Eq. (15), we consider a simple configuration consisting of a single biaxial layer covered by an isotropic 2D conductive sheet. In this case, we set $N = 1$ and $I_1 = 0$ to indicate that no 2D sheet is present at the interface between the 3D layer and the substrate. Eq. (15) then contains only the sum over $l_1$ and reduces to

$$\mathcal{F}_1(\mathbf{q}; \omega) = \sum_{l_1=0}^{1} \mathcal{C}(l_1)(\tanh \xi_1)^{l_1} Q_1^{1-l_1} = \mathcal{C}(1) \tanh \xi_1 + \mathcal{C}(0) Q_1 = 0, \tag{16}$$

where

$$\begin{aligned} \mathcal{C}(0) &= \begin{pmatrix} 1 & Q_2 + I_2 \end{pmatrix} \begin{pmatrix} Q_0 \\ 1 \end{pmatrix} = Q_0 + Q_2 + I_2 \\ \mathcal{C}(1) &= \begin{pmatrix} 1 & Q_2 + I_2 \end{pmatrix} \begin{pmatrix} 0 & Q_1^2 \\ 1 & 0 \end{pmatrix} \begin{pmatrix} Q_0 \\ 1 \end{pmatrix} = Q_1^2 + Q_0(Q_2 + I_2) \end{aligned}. \tag{17}$$

Substituting these functions into Eq. (16) and performing straightforward algebra yields

$$q = \frac{\rho}{k_0 d_1}\left[\mathrm{atan}\left(\frac{\varepsilon_0 \rho}{\varepsilon_{1z}}\right) + \mathrm{atan}\left(\frac{\varepsilon_2 + 2i\alpha_{2/1,eff}\,q}{\varepsilon_{1z}}\rho\right)\right],$$

$$\rho = i\sqrt{\frac{\varepsilon_{1z}}{\varepsilon_{1x}\cos^2\varphi + \varepsilon_{1y}\sin^2\varphi}}.$$

(18)

Equation (18) reproduces the previously reported dispersion relation of polaritons supported by a single finite-thickness biaxial layer capped by a 2D conductive sheet [27]. Similarly, an explicit expression for a twisted bilayer ($N = 2$) with a single isotropic 2D conductive sheet between the layers is obtained from Eq. (15) by setting $I_1 = I_3 = 0$. One finds

$$\mathcal{F}_2(\mathbf{q};\omega) \equiv \sum_{l_1=0}^{1}\sum_{l_2=0}^{1} \mathcal{C}(l_1,l_2)(\tanh\xi_1)^{l_1}(\tanh\xi_2)^{l_2} Q_1^{1-l_1} Q_2^{1-l_2} = 0, \quad (19)$$

where

$$\mathcal{C}(l_1,l_2) = \begin{pmatrix} 1 & Q_3 \end{pmatrix}\begin{pmatrix} (1-l_2) & Q_2^2 l_2 + I_2(1-l_2) \\ l_2 & I_2 l_2 + (1-l_2) \end{pmatrix}\begin{pmatrix} (1-l_1) & Q_1^2 l_1 \\ l_1 & (1-l_1) \end{pmatrix}\begin{pmatrix} Q_0 \\ 1 \end{pmatrix}. \quad (20)$$

Expanding Eq. (19) yields the compact form

$$\mathcal{C}(1,1)\tanh\xi_1 \tanh\xi_2 + \mathcal{C}(0,1)\tanh\xi_2\, Q_1 + \mathcal{C}(1,0)\tanh\xi_1\, Q_2 + \mathcal{C}(0,0)Q_1 Q_2 = 0, \quad (21)$$

where the coeficients $\mathcal{C}(l_1,l_2)$ obtained from Eq. (20) are

$$\begin{aligned}
\mathcal{C}(1,1) &= Q_1^2 Q_3 + Q_0 Q_2^2 + Q_0 Q_3 I_2 \\
\mathcal{C}(0,1) &= Q_3(Q_0 + I_2) + Q_2^2 \\
\mathcal{C}(1,0) &= Q_1^2 + Q_0(Q_3 + I_2) \\
\mathcal{C}(0,0) &= Q_0 + Q_3 + I_2
\end{aligned}. \quad (22)$$

Additional examples of Eq. (15) applied to twisted bilayers and trilayers with interfacial 2D conductive sheets are provided in Ref. [48].

Solving $\mathcal{F}_N(\mathbf{q};\omega) = 0$ is not straightforward because the in-plane polaritonic wavevector $\mathbf{q}$ is complex. We therefore proceed as follows: for each real momentum component $\mathbf{q}_r = \mathrm{Re}(\mathbf{q})$, we seek the value of $\mathbf{q}_i$ such that $\mathbf{q} = \mathbf{q}_r + i\mathbf{q}_i$ minimizes $|\mathcal{F}_N(\mathbf{q};\omega)|$, and plot $-\log_{10}(|\mathcal{F}_N(\mathbf{q};\omega)|)$ as a function of $\mathbf{q}_r$. This quantity measures the order-of-magnitude proximity to a root; for example, $-\log_{10}(|\mathcal{F}_N(\mathbf{q};\omega)|) = 3$ corresponds to $|\mathcal{F}_N(\mathbf{q};\omega)| = 10^{-3}$. The resulting bright maxima therefore indicate regions where the dispersion condition is satisfied with high accuracy, and the roots can be extracted to any desired numerical precision through subsequent local refinement. Note that a more rigorous solution to obtain the precise direction of $\mathbf{q}_i$ can be performed [50,51], which is out of scope of this work.

We next apply this procedure to benchmark the analytical results against well-established numerical methods by considering the solutions of $\mathcal{F}_N(\mathbf{q};\omega) = 0$ for a twisted α-MoO$_3$ bilayer (Fig. 2a), obtained by setting $I_2 = 0$ in Eq. (7). Both α-MoO$_3$ layers have the same thickness, $d_1 = d_2 = 200$ nm, which is typical of exfoliated flakes used in twistoptics [24, 25]. The bilayer is bounded to a semi-infinite air superstrate and a semi-infinite SiO$_2$ substrate. The two layers are rotated relative to each other by $\theta_{1,2} = 63°$, a configuration known to enable PhP canalization near $\omega = 910$ cm$^{-1}$ along the in-plane direction indicated by $\mathbf{r}_c$ in Fig. 2a [24]. This direction forms an angle $\varphi_c = 25°$ with respect to the [100] crystal axis of the top α-MoO$_3$ layer. A top view of the heterostructure is shown in Fig. 2b.

With the geometry thus defined, we first analyze the polaritonic dispersion along the canalization direction $\mathbf{r}_c$. Specifically, we solve Eq. (15) in polar coordinates at the fixed angle $\varphi_c$; i.e, $\mathcal{F}(\mathbf{q} = (q,\varphi_c);\omega) = 0$, while sweeping the frequency from $\omega = 870$ cm$^{-1}$ to $\omega = 950$ cm$^{-1}$. The colormap in Fig. 2c displays $-\log_{10}(|\mathcal{F}((q, \varphi_c); \omega)|)$ as a function of $q_r$ and $\omega$, revealing multiple dispersion branches as bright maxima (highlighted by white curves). These branches correspond to distinct polaritonic modes: the branch with the smallest $q_r$ is typically the fundamental mode, followed by higher-order modes. The extracted radial component of the complex wavevector $q = q_r + iq_i$ can then be used to quantify modal propagation along $\mathbf{r}_c$; for instance, the polariton wavelength is $\lambda = 2\pi(q_r k_0)^{-1}$, the propagation length is $L_p = (q_i k_0)^{-1}$, and the group follows from $\mathbf{v}_{gr} = \nabla_{\mathbf{K}}\omega$.

Furthermore, we analyze the angular dependence of the PhP dispersion by fixing the frequency and varying the in-plane polar angle $\varphi$, thereby reconstructing the isofrequency curves (IFCs). Specifically, we consider $\omega = 890$ cm$^{-1}$ and $\omega = 910$ cm$^{-1}$ and solve $\mathcal{F}(\mathbf{q}; \omega) = 0$ while sweeping $\varphi$. The bright maxima in Figs. 2d,e trace the IFCs of the supported modes (again highlighted by white curves). The pronounced asymmetry of these IFCs reflects strongly direction-dependent propagation in the twisted bilayer. Notably, at $\omega = 910$ cm$^{-1}$, two nearly flat IFC branches emerge. Since the polariton energy flow is approximately normal to the IFC, these nearly flat branches imply strongly collimated, nearly diffraction-free propagation, corresponding to canalized PhPs, in agreement with previous observations [21-24].

Finally, we compare the results in Figs. 2c-e with the reflection coefficient computed using the TMM, shown in Figs. 2i-k [26], finding an excellent agreement between both calculations.

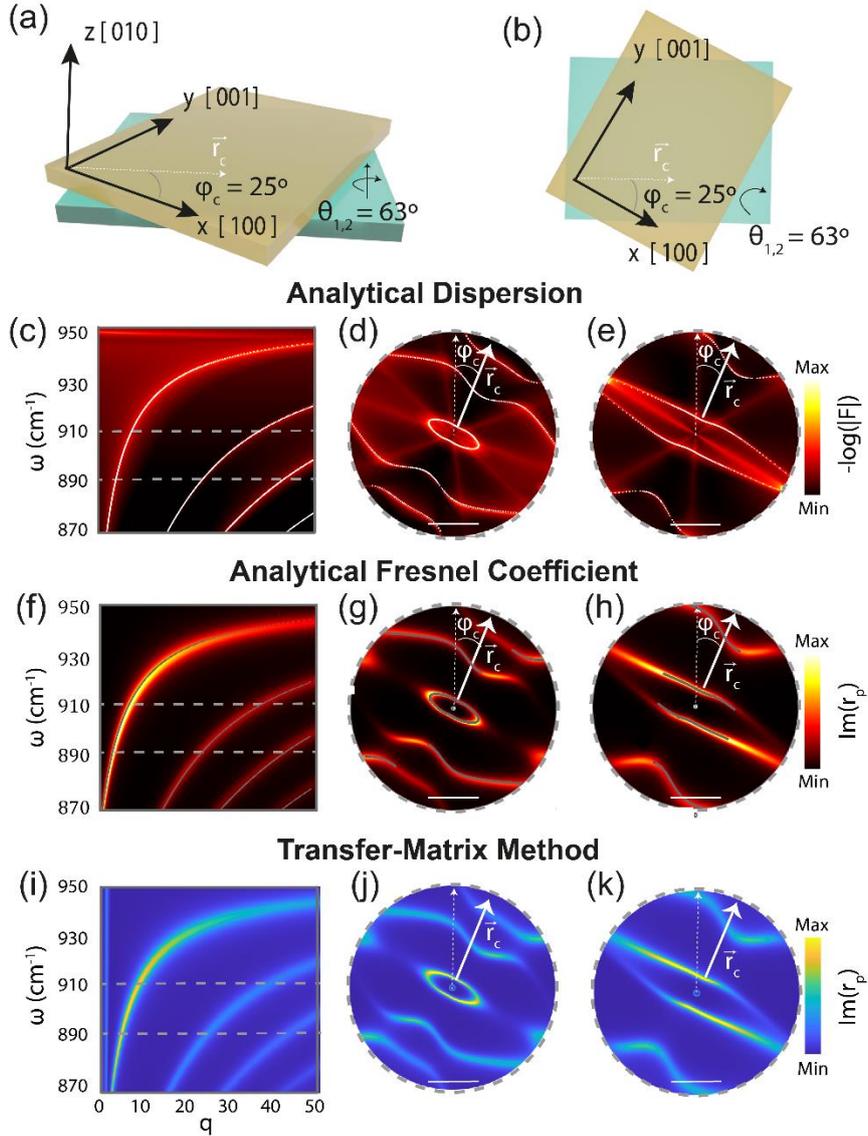

FIG. 2. Validation of the high-momentum approximation in a twisted α-MoO$_3$ bilayer. a) 3D schematics of the heterostructure. The coordinate system is aligned with the crystallographic axes of the top layer [100], [001] and [010].

Both layers have thicknesses $d_1 = d_2 = 200$ nm, and the bottom layer is rotated by $\theta_{1,2} = 63°$ with respect to the top layer. The canalization direction is indicated by $\mathbf{r}_c$, defined by the in-plane angle $\varphi_c = 25°$. The superstrate is air and the substrate is SiO$_2$. b) Top view of a). c) Polaritonic dispersion along $\mathbf{r}_c$ obtained from the high-momentum dispersion equation (Eq. (15); bilayer case), plotted as $-\log_{10}(|\mathcal{F}(\mathbf{q};\omega)|)$ versus real momentum $q_r$ and frequency $\omega$ (within RB2). Bright ridges mark the roots $\mathcal{F} = 0$. Dashed horizontal lines indicate $\omega = 890$ cm$^{-1}$ and $\omega = 910$ cm$^{-1}$. d,e) IFC maps at $\omega = 890$ cm$^{-1}$ and $\omega = 910$ cm$^{-1}$, respectively, computed as $-\log_{10}(|\mathcal{F}(\mathbf{q};\omega)|)$; $\varphi_c$ and $\mathbf{r}_c$ are indicated. f) Same as c), but obtained from the analytical out-of-plane Fresnel reflection coefficient (Eq. (29)) and plotted as $\mathrm{Im}(r_p)$. Dashed lines indicate $\omega = 890$ cm$^{-1}$ and $\omega = 910$ cm$^{-1}$. g,h) IFCs at $\omega = 890$ cm$^{-1}$ and $\omega = 910$ cm$^{-1}$, respectively, obtained from Eq. (29); gray curves (when shown) indicate the solutions extracted from the dispersion equation in Eq. (23). $\varphi_c$ and $\mathbf{r}_c$ describe the direction depicted in a) and b). h) Same as g) calculated at $\omega = 910$ cm$^{-1}$. i-k) TMM reference calculations: dispersion map i) and IFC maps at $= 890$ cm$^{-1}$ j) and $\omega = 910$ cm$^{-1}$ k), plotted as $\mathrm{Im}(r_p)$. The scale bar in panels (d,e, g, h, j, k) corresponds to $q = 25$.

To enable the calculation of polaritonic fields under more realistic conditions, we further develop our analytical approach by imposing that the magnitude of the upgoing field amplitude at the top boundary ($a^\uparrow_{N+1}$) is at its maximum (as typically assumed in near-field excitation/detection schemes) and then solve the linear system in Eq. (4). Restricting the analysis to the extraordinary components and starting from the bottom boundary condition (interface 1/0), we iteratively solve the resulting set of equations, as detailed in Ref. [47]. Within this formalism, the polariton dispersion can be written as the implicit equation:

$$q_l = \frac{\rho_N}{k_0 d_N}\left[\mathrm{atan}\left(\frac{\left(\varepsilon_{N+1} + \frac{2I_{N+1/N,eff}}{q}\right)\rho_N}{\varepsilon_{N,z}}\right) + \mathrm{atan}\left(\frac{\rho_N}{D_{N-1}(q)\varepsilon_{N,z}q}\right) + \pi l\right], \quad (23)$$

with the recursive relation

$$D_j(q_l) = \frac{D_{j-1}(q_l)\Sigma_j\cos(\xi_{j,z}) + i\sin(\xi_{j,z})}{\Sigma_j\left(1 + 2D_{j-1}(q_l)I_j\right)\cos(\xi_{j,z}) + \left[iD_{j-1}(q)\Sigma_j^2 + 2iI_j\right]\sin(\xi_{j,z})}, \quad (24)$$

for $j = 0, \ldots, N-1$, where $\rho_N = \pm i\frac{q_l}{q_{N,z}}$. Here the subscript $N$ refers to the properties of the top finite-thickness layer, and the index $l$ labels the l-th solution of Eq. (23).

In addition to the dispersion obtained from Eqs. (23), our formalism also provides the complex electric-field amplitudes in each layer of the twisted heterostructure [48]. In particular, for the j-th layer ($j = 1, \ldots, N$), the upgoing and downgoing plane-wave amplitudes can be written as

$$a^\uparrow_j = \frac{E_{j+1}e^{iq_{j+1,z}k_0\sum_{i=1}^j d_i} + E_{j+1}F_{j+1}e^{-iq_{j+1,z}k_0\sum_{i=1}^j d_i}}{e^{iq_{j,z}k_0\sum_{i=1}^j d_i} + \left(\frac{D_{j-1}\Sigma_j - 1}{D_{j-1}\Sigma_j + 1}\right)e^{-iq_{j,z}k_0(2d_j - \sum_{i=1}^j d_i)}} a^\uparrow_{j+1} = E_j a^\uparrow_{N+1}, \quad (25)$$

$$a^\downarrow_j = e^{2iq_{j,z}k_0\sum_{i=1}^{j-1} d_i}\left(\frac{D_{j-1}\Sigma_j - 1}{D_{j-1}\Sigma_j + 1}\right)a^\uparrow_j = F_j E_j a^\uparrow_{N+1}, \quad (26)$$

Here we use $q_{N+1,z} = iq$ with $E_{N+1} = 1$ and $F_{N+1} = 0$. In the substrate, only a single (downgoing) amplitude $a^\downarrow_0$ is required, which satisfies

$$a^\downarrow_0 = E_1(1 + F_1)a^\uparrow_{N+1} \quad (27)$$

Eqs. (25) enable the reconstruction of the spatial electric field distribution of the l-th mode supported by the heterostructure, providing direct real-space visualization of the corresponding polaritonic fields. Additionally, our approach provides the $p$-polarized Fresnel reflection coefficient corresponding to the out-of-plane electric field component of a twisted planar heterostructure, defined as $r_p = \frac{E_{N+1,r,z}}{E_{N+1,i,z}}$. To obtain $r_p$, we set for both upgoing (reflected) and downgoing (incident) $p$-polarized waves in the superstrate ($N + 1$) and enforce the boundary conditions at the top interface $z = \sum_{i=1}^{N} d_i$. Specifically, this amounts to replacing the superstrate field in Eq. (8) by (see Ref. [48]):

$$\mathbf{E}_{N+1} = a_{N+1}^{\uparrow} \mathbf{p}_{N+1,+} e^{i\xi_{N+1}} \rightarrow a_{N+1}^{\uparrow} \mathbf{p}_{N+1,+} e^{i\xi_{N+1}} + a_{N+1}^{\downarrow} \mathbf{p}_{N+1,-} e^{-i\xi_{N+1}}. \quad (28)$$

With this modification, the $p$-polarized Fresnel reflection coefficient of the heterostructure becomes:

$$r_p(q) = \frac{(D_N(q)\varepsilon_S q - 1)}{(D_N(q)\varepsilon_S q + 1)}, \quad (29)$$

where $D_N(q)$ is obtained from the recursion in Eq. (24) (with the appropriate termination at the top interface). Equation (29) can be used to analyze both the polaritonic and the IFCs of an arbitrary twisted heterostructure by examining the poles (or similarly, the maxima of $\text{Im}(r_p)$) as a function of momentum and frequency.

We further validate our results for the stack in Fig. 2a by computing the imaginary part of the analytical Fresnel reflection coefficient in Eq. (29) (Figs. 2f-h). The excellent agreement with the corresponding TMM calculations (Figs. 2i-k) confirms the validity of Eq. (29). In addition to the colormap representations in Figs. 2f-h, we overlay the solutions of Eq. (23) as solid gray curves. Importantly, unlike Eq. **¡Error! No se encuentra el origen de la referencia.**, Eq. (23) yields each polaritonic branch separately by selecting the mode index $l$ (i.e., $l = 0,1,2$ in Fig. 2g and $l = 0,1$ in Fig. 2h).

To predict the polaritonic fields in real space, a widely used approach is to model the excitation source as a point dipole. The electric field radiated by a dipole can be written as $\mathbf{E}(\mathbf{r}) = \hat{G}(\mathbf{r}, \mathbf{r}')\mathbf{p}(\mathbf{r}')$, where $\mathbf{p}(\mathbf{r}')$ is the dipole moment and $\hat{G}(\mathbf{r}, \mathbf{r}')$ is the Dyadic Green's Function (DGF). Within this framework, the polaritonic fields can be obtained from the reflected contribution of the Green's function associated with the multilayer stack. Following Refs. [30,31], we evaluate the $p$-polarized out-of-plane component of the reflected DGF as

$$\hat{G}_R(\mathbf{r}_\parallel, z; z') \rightarrow \hat{G}_{zz}(\mathbf{r}_\parallel, z; z') = -\frac{k_0}{8\pi^2} \iint d\mathbf{q}\, q\, r_p\, e^{i\mathbf{k}_\parallel \cdot \mathbf{r}_\parallel} e^{ik_z(z+z')}. \quad (30)$$

Equation **¡Error! No se encuentra el origen de la referencia.** gives the out-of-plane electric-field component generated by a vertical point dipole $\mathbf{p} = (0 \quad 0 \quad 1)^T$. The Fresnel coefficient $r_p$, encodes the material properties and thus constitutes an important ingredient for the polaritonic response of a heterostructure. In practice, Eq. **¡Error! No se encuentra el origen de la referencia.** can be evaluated numerically, as we show below.

The polaritons excited by a vertical point dipole source placed above the heterostructure of Figs. 2a,b is shown in Fig. 3a for an excitation frequency $\omega = 910$ cm$^{-1}$. The dipole is placed at $z' = 100$ nm above the top biaxial layer. Its electric field presents a continuum of plane waves with a broad in-plane momentum spectrum, so that highly confined polaritons can be excited. The resulting canalized polaritons are clearly visible as alternating red/blue fringes propagating along the $\mathbf{r}_c$ direction of the heterostructure. This analytical result closely matches the corresponding full-wave simulations (COMSOL Multiphysics) in Fig. 3b, confirming that Eq. (30) provides an efficient route to compute real-space in-plane field distributions of strongly confined polaritons. In addition, the field-amplitude coefficients defined in Eqs. (25) enable an entirely analytical reconstruction of cross-sectional field profiles within the heterostructure. Figs. 3c,e show the real parts of the in-plane and out-of-plane electric-field components, $\text{Re}(\mathbf{E}_\parallel)$ and $\text{Re}(\mathbf{E}_z)$, respectively, in the $\mathbf{r}_c - z$ plane for the fundamental ($l = 0$) mode. We obtain again an excellent agreement with numerical simulations ($\text{Re}(\mathbf{E}_\parallel)$ in Fig. 3d and $\text{Re}(\mathbf{E}_z)$ in Fig. 3f), further supporting the validity of Eqs. (25).

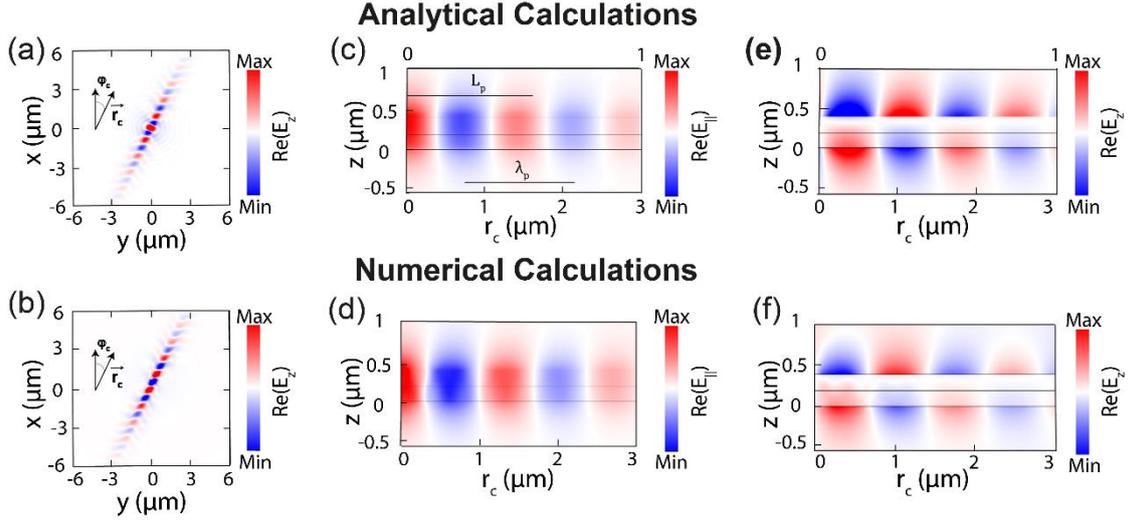

FIG. 3. Real-space validation of the high-momentum approximation for a twisted α-MoO$_3$ bilayer at the "magic" angle $\theta_{1,2} = 63°$ and $\omega = 910$ cm$^{-1}$. a) Semi-analytical real-space map of canalized polaritons computed from the reflected dyadic Green´s function (Eq. (30)) for a vertical point dipole placed above the heterostructure. The arrow $\mathbf{r}_c$ indicates the canalization direction, consistent with the dispersion analysis in Fig 2. b) Corresponding full-wave simulation of the same configuration, showing excellent agreement with a). c, e) Analytical cross-sections of the fundamental ($l = 0$) polaritonic mode in the $\mathbf{r}_c - \mathbf{z}$ plane obtained from the field amplitudes in Eqs. (25-27): c) Re($\mathbf{E}_\parallel$) and e) Re($\mathbf{E}_z$). Black horizontal lines mark the interfaces $N_{3/2}$, $N_{2/1}$ and $N_{1/0}$. In c), the polariton wavelength $\lambda_p$ and the propagation length $L_p$ are indicated. d, f) Full-wave (quasi-eigenmode) simulations of the cross-sections of the fundamental mode in the same the $\mathbf{r}_c - \mathbf{z}$ plane: d) Re($\mathbf{E}_\parallel$) and f) Re($\mathbf{E}_z$).

We now apply our analytical framework to study the propagation of polaritons in hybrid twisted heterostructures that combine multiple anisotropic vdW crystals with 2D conductive sheets. As a representative example (Fig. 4), we study a two-layer stack consisting of a twisted α-MoO$_3$/α-V$_2$O$_5$ bilayer capped with a graphene sheet. The layer thicknesses, $d_1 = 200$ nm for α-V$_2$O$_5$ and $d_2 = 100$ nm for α-MoO$_3$, together with the relative twist angle $\theta_{1,2} = -113°$ and the operating frequency $\omega = 640$ cm$^{-1}$, are chosen to yield PhP canalization in the bilayer. The graphene response is modeled through its optical conductivity within the random phase approximation [52], assuming a Fermi energy $E_F = 0.3$ eV, such that the sheet supports graphene plasmon polaritons (GPPs) in the mid-IR [53]. The whole heterostructure is embedded between air (as superstrate) and SiO$_2$ (as substrate).

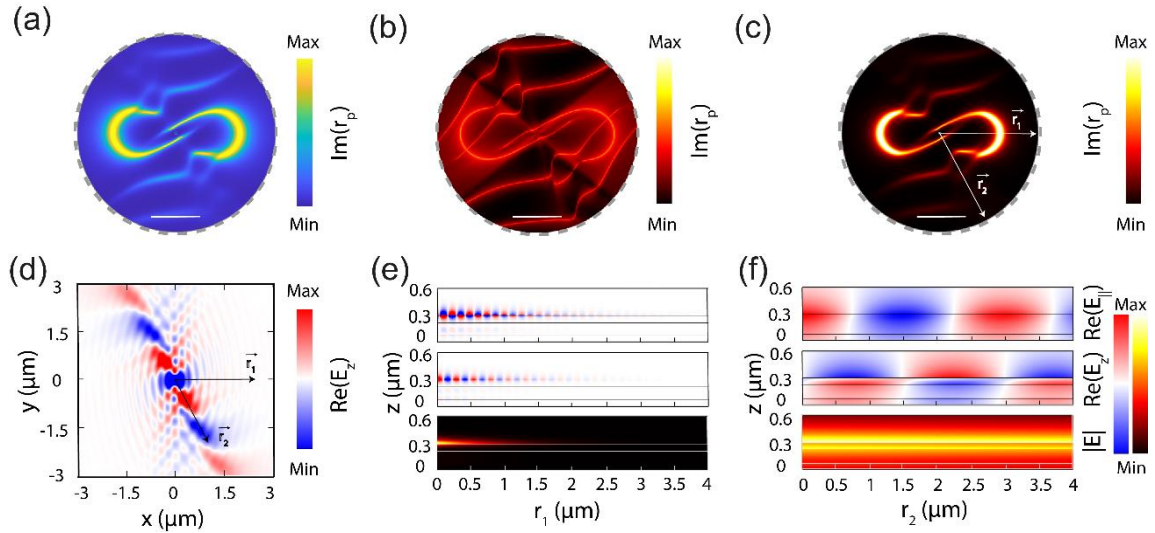

FIG. 4. Fourier- and real-space characterization of hybrid GPs/HPhPs in graphene/α-MoO$_3$/α-V$_2$O$_5$. The heterostructure consists of a graphene sheet on top of a twisted α-MoO$_3$ ($d_2 = 100$ nm)/α-V$_2$O$_5$ ($d_1 = 200$ nm) bilayer with relative twist angle $\theta_{1,2} = -113°$, evaluated at $\omega = 640$ cm$^{-1}$. a) IFC computed numerically using the TMM. b) Analytical IFC obtained from the high-momentum dispersion equation (Eq. (15)). c) Analytical IFC extracted from the out-of-plane Fresnel reflection coefficient (Eq. (29)). The scale bar in a-c) corresponds to $q = 50$. d) Real-space polaritonic electric-field distribution computed from the reflected dyadic Green´s function (Eq. (30)). e) Cross-sectional profiles of the fundamental mode along direction $\mathbf{r}_1$ (defined in panel c)): Re($\mathbf{E}_\parallel$) (top), Re($\mathbf{E}_z$) (middle), and |$\mathbf{E}$| (bottom). f) Same as e), but along direction $\mathbf{r}_2$.

We compare the IFC computed with the TMM (Fig. 4a) with the IFCs obtained from Eq. **¡Error! No se encuentra el origen de la referencia.** and Eq. (29) (Fig. 4b and Fig. 4c, respectively). In both cases we find an excellent agreement between the two methods with Eq. (29) providing the closest match to TMM, consistent with the trend observed in Fig. 2. To clarify how these IFC features translate into real-space propagation, we perform analytical calculations in Figs. 4d-f. The in-plane field distribution obtained from Eq. (29) is shown in Fig. 4d. Two distinct behaviors are apparent: i) an elliptical propagation pattern associated with the brightest regions of the IFC in Fig. 4c, and ii) a canalized propagation component. We further use Eqs. (25-27) to compute cross-sectional electric-field profiles of the fundamental mode along the two in-plane directions, $\mathbf{r}_1$ and $\mathbf{r}_2$, indicated in Fig. 4c. Fig. 4e reports Re($\mathbf{E}_\parallel$) (top), Re($\mathbf{E}_z$) (middle), and |$\mathbf{E}$| (bottom) along $\mathbf{r}_1$. The fields are strongly confined to the top interface, revealing a surface-like mode that we attribute to hybridization between GPs and the HPhPs. The corresponding wavelength along $\mathbf{r}_1$ is $\lambda_p \sim 260$ nm. In contrast, along $\mathbf{r}_2$ (Fig. 4f) the field profiles exhibit a predominantly volume-like character in Re($\mathbf{E}_\parallel$) (top), Re($\mathbf{E}_z$) (middle), and |$\mathbf{E}$| (bottom), together with a much larger wavelength, $\lambda_p \sim 3.1$ μm.

## IV. THIN-FILM APPROXIMATION IN ARBITRARILY LARGE HETEROSTRUCTURES

When polaritons are not strongly confined, the high-momentum approximation becomes less accurate and may fail. In this regime, particularly when the constituent layers in a heterostructure are sufficiently thin such that $k_z d \ll 1$, the thin-film approximation provides a convenient alternative. In this approximation, each finite-thickness 3D layer is replaced by an infinitesimally thin effective conductive sheet. This yields compact analytical expressions and a substantially reduced computational cost, which is advantageous for rapid exploration and optimization of large multilayer stacks.

However, we note that the thin-film approximation implicitly assumes a small phase accumulation across each layer. In particular, the propagation phases $\xi_{j,\gamma}$ in Eq. (7) must remain sufficiently small, which depends on the layer thickness, operating frequency, and the out-of-plane wavevector. When these phases are not negligible, the approximation can deviate from experimental observations and should be benchmarked against numerical approaches (i.e., TMM) for the specific parameter range of interest [24,25].

Under the thin-film approximation, each (sufficiently thin) biaxial 3D layer can be replaced by an equivalent anisotropic sheet characterized by an effective conductivity tensor [49]:

$$\hat{\sigma}_j = \frac{c d_j}{2 i \lambda} \hat{\varepsilon}_j. \tag{31}$$

To implement this approximation, we take the limit $k_0 d_j \to 0$ for all 3D layers in Eq. (4) (see Ref. [47] and Ref. [48]). In this limit, and after applying Gaussian elimination, the determinant of the matrix in Eq. (5) reduces to that of a $2 \times 2$ matrix,

$$|M| = \begin{vmatrix} 2\mathbf{s}\hat{\alpha}_{eff}\mathbf{s} + q_{N+1,z} + q_{0,z} & 2\mathbf{s}\hat{\alpha}_{eff}\mathbf{p} \\ 2\mathbf{p}\hat{\alpha}_{eff}\mathbf{s} & 2\mathbf{p}\hat{\alpha}_{eff}\mathbf{p} + \frac{\varepsilon_{N+1}}{q_{N+1,z}} + \frac{\varepsilon_0}{q_{0,z}} \end{vmatrix}, \tag{32}$$

Where **s** and **p** denote the usual $s$- and $p$-polarization basis vectors, and

$$\hat{\alpha}_{eff} = \sum_j \hat{\alpha}_j = \begin{pmatrix} \alpha_{eff,xx} & \alpha_{eff,xy} \\ \alpha_{eff,xy} & \alpha_{eff,yy} \end{pmatrix}, \quad (33)$$

is the effective (normalized) conductivity tensor obtained by summing the contributions of all 2D sheets in the heterostructure, consistently with Ohm's law.

Expressed in terms of the in-plane polar angle $\varphi$ and the twist angle $\phi_j$ of the j-th sheet (referenced to the principal axes of the bottom one), the components of $\alpha_{eff}$ are

$$\alpha_{eff,xx} = \sum_j [\alpha_{j,xx} \cos^2(\varphi + \phi_j) + \alpha_{j,yy} \sin^2(\varphi + \phi_j)]$$
$$\alpha_{eff,yy} = \sum_j [\alpha_{j,xx} \sin^2(\varphi + \phi_j) + \alpha_{j,yy} \cos^2(\varphi + \phi_j)], \quad (34)$$
$$\alpha_{eff,xy} = \sum_j [(\alpha_{j,yy} - \alpha_{j,xx}) \sin(\varphi + \phi_j) \cos(\varphi + \phi_j)]$$

The thin-film dispersion relation follows from $|M| = 0$ in Eq. (32), which can be written explicitly as the algebraic equation

$$\left[ \alpha_{eff,xx} q_y^2 + \alpha_{eff,yy} q_x^2 - 2\alpha_{eff,xy} q_x q_y + \frac{q^2}{2}(iq_{N+1,z} + iq_{0,z}) \right]$$
$$\left[ \alpha_{eff,xx} q_x^2 + \alpha_{eff,yy} q_y^2 + 2\alpha_{eff,xy} q_x q_y + \frac{q^2}{2}\left(\frac{\varepsilon_{N+1}}{iq_{N+1,z}} + \frac{\varepsilon_0}{iq_{0,z}}\right) \right] = \quad (35)$$
$$[(\alpha_{eff,yy} - \alpha_{eff,xx}) q_x q_y + \alpha_{eff,xy}(q_x^2 - q_y^2)]^2,$$

where the momentum dependence of Eq. (35) enters through $q_{0,z}$ and $q_{N+1,z}$. Eq. (35) is the natural generalization of the thin-film dispersion previously derived for a single 2D conductive sheet [47], now featuring an effective normalized, generally non-diagonal conductivity tensor $\hat{\alpha}_{eff}$. As illustrated in Fig. 5a, within this approximation the entire multilayer heterostructure reduces to a single effective 2D sheet embedded between two semi-infinite isotropic media.

Two limiting cases of Eq. (35) are particularly instructive. Following Ref. [50], in the high-enough momenta and sufficiently small tensor components $|\alpha_{eff,ij}| \ll 1$, the first three terms in the first bracket on the left-hand side of Eq. (35) can be neglected, yielding

$$\alpha_{eff,xx} q_x^2 + \alpha_{eff,yy} q_y^2 + 2\alpha_{eff,xy} q_x q_y + \frac{\varepsilon_{N+1} + \varepsilon_0}{2i} =$$
$$\frac{2[(\alpha_{eff,yy} - \alpha_{eff,xx}) q_x q_y + \alpha_{eff,xy}(q_x^2 - q_y^2)]^2}{q^3}. \quad (36)$$

In Eq. (36), the left-hand side scales as $q^2$, whereas the right-hand side scales as $q$. Therefore, to leading order the right-hand side can be neglected, and Eq. (36) reduces to an explicit expression for the polaritonic momentum:

$$q = \frac{i(\varepsilon_{N+1} + \varepsilon_0)}{2(\alpha_{eff,xx} \cos^2\varphi + \alpha_{eff,yy} \sin^2\varphi + 2\alpha_{eff,xy} \sin\varphi \cos\varphi)}. \quad (37)$$

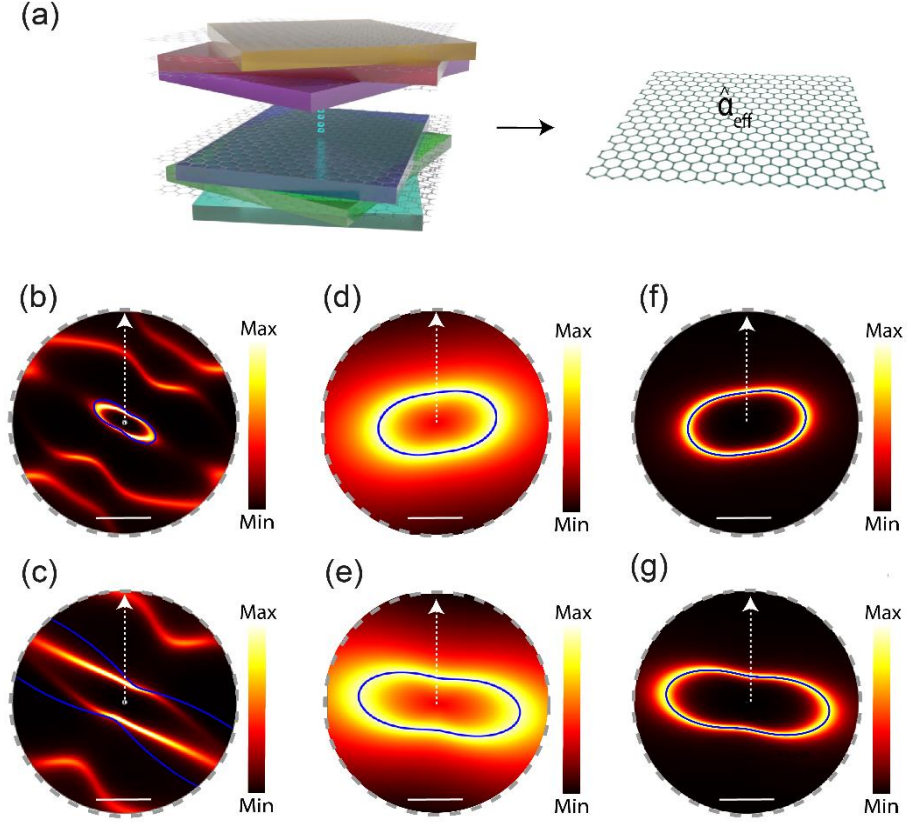

FIG. 5. Range of validity of the thin-film approximation for twisted heterostructures. a) Conceptual schematics of the thin-film approximation. The full multilayer is replaced by a single effective 2D conductive sheet characterized by the normalized tensor $\hat{\alpha}_{eff}$, embedded between two semi-infinite isotropic media. b,c) IFCs of the twisted α-MoO₃ bilayer of Fig. 2a at $\omega = 890$ cm⁻¹ and $\omega = 910$ cm⁻¹, respectively. The color plots are obtained from the out-of-plane Fresnel reflection coefficient in the high-momentum approximation (Eq. (29)), while the blue curves correspond to the explicit thin-film expression (Eq. (37)). The agreement is still reasonable at $\omega = 890$ cm⁻¹, whereas at $\omega = 910$ cm⁻¹ the thin-film limit fails to reproduce the expected IFC features. The white dashed arrow marks the [100] axis of the top α-MoO₃ layer. d,e) IFCs at $\omega = 2000$ cm⁻¹ for a homostructure of three rotated BP monolayers embedded in air with twist angles $\theta_{1,2} = 100°$ and $\theta_{1,3} = 70°$ (d) and $\theta_{1,2} = 80°$ and $\theta_{1,3} = 70°$ (e). The colormaps are computed from the thin-film dispersion equation (Eq. (35)), and blue curves from Eq. (37). f,g) Same as d,e), but the colormaps are obtained from the high-momentum Fresnel coefficient (Eq. (29)). In this BP homostructure the accessible momenta are sufficiently large such that Eqs. (29), (35), and (37) yield essentially identical IFCs. The scale bars are $q = 25$ in b,c) and $q = 2500$ in d-g).

Equation (37) provides the dispersion of TM modes supported by an effective 2D conductive sheet characterized by the tensor $\hat{\alpha}_{eff}$ defined in Eq. (33). It is therefore also applicable in the thin-film approximation and in the high-momentum regime.

To assess the practical validity of Eq. (35) and Eq. (37), we performed calculations analogous to those presented in Fig. 2. We first benchmark the thin-film approximation against the high-momentum formalism developed in Section II. Figs. 5b,c show the same IFCs as in Figs. 2d,e at $\omega = 890$ cm⁻¹ and $\omega = 910$ cm⁻¹, respectively, computed from the high-momentum Fresnel coefficient (Eq. (29); colormaps). On top of these maps we overlay the solutions of Eq. (37) for the same system (blue curves). While a partial agreement is still observed at $\omega = 890$ cm⁻¹ (Fig. 5b), Eq. (37) fails to reproduce the IFC features at $\omega = 910$ cm⁻¹ (Fig. 5c). Moreover, Eq. (37) does not carry mode-index dependence, and therefore it cannot capture higher-order branches. Overall, these results indicate that the thin-film approximation is generally inaccurate for the α-MoO₃ bilayer considered here, consistent with previous reports [24,25].

Nevertheless, there are relevant scenarios in which the thin-film approximation becomes reliable. To illustrate this, we consider a homostructure composed of three rotated black-phosphorus (BP) monolayers. The normalized BP conductivity is taken from Ref. [54]; in doing so, we neglect possible interlayer interaction effects that are not captured by the single-layer conductivity model. Figs. 5d,e compare Eq. (35) (colormaps) and Eq. (37) (blue curves) at $\omega = 2000$ cm$^{-1}$ for two sets of twist angles: $\theta_{1,2} = 100°$ and $\theta_{1,3} = 70°$ (Fig. 5d) and $\theta_{1,2} = 80°$ and $\theta_{1,3} = 70°$ (Fig. 5e). The close agreement confirms that, for this system, Eq. (37) provides an accurate high-momentum limit of Eq. (35).

Importantly, the validity of the thin-film description must ultimately be assessed against a more complete description. This is done in Figs. 5f,g, where the IFCs obtained from the high-momentum Fresnel coefficient (Eq. (29); colormaps) are compared to the explicit thin-film expression (Eq. (37); blue curves). The agreement indicates that, for this BP homostructure, the thin-film approximation is consistent with the high-momentum treatment. More broadly, these results emphasize that adopting the thin-film limit should be conditioned on benchmarking Eq. (37) against Eq. (29) (when strong confinement applies) or, in general, against full numerical methods.

## V. CONCLUSIONS

In summary, we have developed a general analytical framework to characterize polaritons in twisted planar heterostructures composed of anisotropic 3D thin layers and 2D conductive sheets. Our formalism provides compact expressions to compute, in arbitrarily large stacks, the polaritonic dispersion and IFCs, vectorial electric-field distributions, out-of-plane Fresnel reflection coefficients, and the reflected dyadic Green's function enabling real-space propagation maps. We have tested our framework against established numerical approaches -including full-wave real-space simulations- finding an excellent agreement and thereby validating the theory.

Beyond offering an efficient alternative to purely numerical modeling, our results establish a practical theoretical foundation for Twistoptics: they enable *a priori* prediction and rapid optimization of twist-engineered polaritonic propagation -such as canalization and other highly anisotropic transport regimes- across a wide range of materials and spectral domains. The availability of fast analytical calculations is particularly valuable for exploring high-dimensional parameter spaces (twist angles, thicknesses, conductivities, and frequency), and it naturally lends itself to generating large training datasets for data-driven approaches. In this sense, the proposed framework can directly support machine-learning strategies for inverse design of twisted heterostructures with targeted polaritonic functionalities.

We also introduced a thin-film approximation that can serve as a complementary, low-cost description in suitable parameter regimes, provided it is carefully benchmarked against more complete methods. Finally, we provide an open-access numerical implementation of the framework [55], enabling efficient exploration and design of polaritonic properties in complex twisted heterostructures comprising an arbitrary number of finite-thickness anisotropic layers and interfacial 2D conductive sheets with arbitrary twist angles.

**Funding**

The authors acknowledge the Spanish Ministry of Science and Innovation (State Plan for Scientific and Technical Research and Innovation grant numbers PID2022-141304NB-I00 and PID2023147676NB-I00). J. Á. C. acknowledges support through the Severo Ochoa program from the Government of the Principality of Asturias (no. PA-22-PF-BP21-100). K.V.V. received the support of a fellowship from "la Caixa" Foundation (ID 100010434) with the fellowship code LCF/BQ/DI21/11860026. G.Á.-P. acknowledges support from the European Union (Marie Skłodowska-Curie Actions, grant agreement No. 101209198). J.M.-S. acknowledges financial support from the Spanish Ministry of Science and Innovation (grant number PID2023- 148457NB-I00 funded by MCIN/AEI/10.13039/501100011033 and FSE +, PCI2022-132953 funded by MCIN/AEI/10.13039/501100011033 and the EU "NextGenerationEU"/PRTR, CNS2024-154342 funded by MICIU/ AEI /10.13039/501100011033"). A.Y.N. acknowledges the Department of

Science, Universities and Innovation of the Basque Government (grant PIBA-2023-1-0007 and the IKUR Strategy). P.A.-G. acknowledges support from the European Research Council under Consolidator grant no. 101044461, TWISTOPTICS and the Spanish Ministry of Science and Innovation (State Plan for Scientific and Technical Research and Innovation grant number PID2019-111156GB-I00). This work was supported by Agencia SEKUENS (Asturias) under grant UONANO IDE/2024/000678 with the support of FEDER funds.

**Notes**

The authors declare no competing financial interest.

**REFERENCES**

[1] A. K. Geim and I. V Grigorieva, Van der Waals heterostructures, Nature **499**, 419 (2013).

[2] K. S. Novoselov, A. Mishchenko, A. Carvalho, and A. H. Castro Neto, 2D materials and van der Waals heterostructures, Science (1979) **353**, aac9439 (2016).

[3] S. Carr, D. Massatt, S. Fang, P. Cazeaux, M. Luskin, and E. Kaxiras, Twistronics: Manipulating the electronic properties of two-dimensional layered structures through their twist angle, Phys Rev B **95**, (2017).

[4] Y. Cao, V. Fatemi, S. Fang, K. Watanabe, T. Taniguchi, E. Kaxiras, and P. Jarillo-Herrero, Unconventional superconductivity in magic-angle graphene superlattices, Nature **556**, 43 (2018).

[5] Y. Cao et al., Correlated insulator behaviour at half-filling in magic-angle graphene superlattices, Nature **556**, 80 (2018).

[6] M. Yankowitz, S. Chen, H. Polshyn, Y. Zhang, K. Watanabe, T. Taniguchi, D. Graf, A. F. Young, and C. R. Dean, Tuning superconductivity in twisted bilayer graphene, Science (1979) **363**, 1059 (2019).

[7] Y. Cao, D. Chowdhury, D. Rodan-Legrain, O. Rubies-Bigorda, K. Watanabe, T. Taniguchi, T. Senthil, and P. Jarillo-Herrero, Strange Metal in Magic-Angle Graphene with near Planckian Dissipation, Phys Rev Lett **124**, (2020).

[8] G. Hu, A. Krasnok, Y. Mazor, C.-W. Qiu, and A. Alù. Moirè Hyperbolic Metasurfaces. Nano Lett. 2020, 20, 5, 3217–3224.

[9] J. Álvarez-Cuervo, M. Obst, S. Dixit *et al.* Unidirectional ray polaritons in twisted asymmetric stacks. Nat Commun **15**, 9042 (2024).

[10] R. Hillenbrand, Y. Abate, M. Liu, X. Chen and D. N. Basov. Visible-to-THz near-field nanoscopy. Nat Rev Mater **10**, 285–310 (2025).

[11] S. Dai et al., Tunable Phonon Polaritons in Atomically Thin van der Waals Crystals of Boron Nitride, Science (1979) **343**, 1125 (2014).

[12] S. Dai et al., Subdiffractional focusing and guiding of polaritonic rays in a natural hyperbolic material, Nat Commun **6**, 6963 (2015).

[13] S. Dai et al., Efficiency of Launching Highly Confined Polaritons by Infrared Light Incident on a Hyperbolic Material, Nano Lett **17**, 5285 (2017).

[14] A. J. Giles et al., Ultralow-loss polaritons in isotopically pure boron nitride, Nat Mater **17**, 134 (2018).

[15] W. Ma et al., In-plane anisotropic and ultra-low-loss polaritons in a natural van der Waals crystal, Nature **562**, 557 (2018).

[16] J. Taboada-Gutiérrez et al., Broad spectral tuning of ultra-low-loss polaritons in a van der Waals crystal by intercalation, Nat Mater **19**, 964 (2020).

[17] D. N. Basov, M. M. Fogler, and F. J. García de Abajo, Polaritons in van der Waals materials, Science (1979) **354**, aag1992 (2016).

[18] T. Low, A. Chaves, J. D. Caldwell, A. Kumar, N. X. Fang, P. Avouris, T. F. Heinz, F. Guinea, L. Martin-Moreno, and F. Koppens, Polaritons in layered two-dimensional materials, Nat Mater **16**, 182 (2017).

[19] P. Li et al., Collective near-field coupling and nonlocal phenomena in infrared-phononic metasurfaces for nano-light canalization, Nat Commun **11**, 3663 (2020).

[20] A. I. F. Tresguerres-Mata et al., Observation of naturally canalized phonon polaritons in LiV2O5 thin layers, Nat Commun **15**, (2024).

[21] G. Hu et al., Topological polaritons and photonic magic angles in twisted α-MoO3 bilayers, Nature **582**, 209 (2020).

[22] M. Chen, X. Lin, T. H. Dinh, Z. Zheng, J. Shen, Q. Ma, H. Chen, P. Jarillo-Herrero, and S. Dai, Configurable phonon polaritons in twisted α-MoO3, Nat Mater **19**, 1307 (2020).

[23] Z. Zheng, F. Sun, W. Huang, J. Jiang, R. Zhan, Y. Ke, H. Chen, and S. Deng, Phonon Polaritons in Twisted Double-Layers of Hyperbolic van der Waals Crystals, Nano Lett. 2020 20, 7, 5301–5308.

[24] J. Duan, N. Capote-Robayna, J. Taboada-Gutiérrez, G. Álvarez-Pérez, I. Prieto, J. Martín-Sánchez, A. Y. Nikitin, and P. Alonso-González, Twisted Nano-Optics: Manipulating Light at the Nanoscale with Twisted Phonon Polaritonic Slabs, Nano Lett **20**, 5323 (2020).

[25] J. Duan et al., Multiple and spectrally robust photonic magic angles in reconfigurable α-MoO3 trilayers, Nat Mater **22**, 867 (2023).

[26] N. C. Passler and A. Paarmann, Generalized 4 × 4 matrix formalism for light propagation in anisotropic stratified media: study of surface phonon polaritons in polar dielectric heterostructures, Journal of the Optical Society of America B **34**, 2128 (2017).

[27] G. Álvarez-Pérez, A. González-Morán, N. Capote-Robayna, K. V. Voronin, J. Duan, V. S. Volkov, P. Alonso-González, and A. Y. Nikitin, Active Tuning of Highly Anisotropic Phonon Polaritons in Van der Waals Crystal Slabs by Gated Graphene, ACS Photonics **9**, 383 (2022).

[28] H. Hu et al., Doping-driven topological polaritons in graphene/α-MoO3 heterostructures, Nat Nanotechnol **17**, 940 (2022).

[29] Z. Zhou et al., Gate-Tuning Hybrid Polaritons in Twisted α-MoO3/Graphene Heterostructures, Nano Lett **23**, 11252 (2023).

[30] A. Y. Nikitin, F. Guinea, F. J. Garcia-Vidal, and L. Martin-Moreno, Fields radiated by a nanoemitter in a graphene sheet, Phys Rev B **84**, 195446 (2011).

[31] A. Y. Nikitin, F. J. Garcia-Vidal, and L. Martin-Moreno, Analytical expressions for the electromagnetic dyadic green's function in graphene and thin layers, IEEE Journal on Selected Topics in Quantum Electronics **19**, (2013).

[32] Y. Hu, J. Álvarez-Cuervo, E. Terán-García, X. Huang, P. Alonso-González. Modulation of Radiative Heat Transfer at the Nanoscale via Topological Polaritons in Twisted van der Waals Crystals. Nanophotonics: e70000.

[33] S. Molesky, Z. Lin, A. Y. Piggott, W. Jin, J. Vucković, and A. W. Rodriguez, Inverse Design in Nanophotonics, Nature Photonics **12**, 659–670 (2018).

[34] M. He, J. Nolen, J. Nordlander, A. Cleri, N. McIlwaine, T. Folland, Y. Tang, B. Landman, J. Maria, and J. Caldwell, Deterministic inverse design of Tamm plasmon thermal emitters with multi-resonant control. Nat. Mater. **20**, 1663–1669 (2021).

[35] J. F. Masson, J. S. Biggins, and E. Ringe, Machine learning for nanoplasmonics. Nat. Nanotechnol. **18**, 111–123 (2023).

[36] R. Claus, R. CLAUS: Polariton Dispersion and Crystal Optics in Monoclinic Materials Polariton Dispersion and Crystal Optics in Monoclinic Materials, 1978.

[37] N. C. Passler et al., Hyperbolic shear polaritons in low-symmetry crystals, Nature **602**, 595 (2022).

[38] G. Hu et al., Real-space nanoimaging of hyperbolic shear polaritons in a monoclinic crystal, Nat Nanotechnol **18**, 64 (2023).

[39] J. Matson et al., Controlling the Propagation Asymmetry of Hyperbolic Shear Polaritons in Beta-Gallium Oxide. Nat Commun **14**, 5240 (2023).

[40] E. Galiffi et al., Extreme Light Confinement and Control in Low-Symmetry Phonon-Polaritonic Crystals, Nat Rev Mater **9**, 9–28 (2024).

[41] G. A. Ermolaev et al., Wandering principal optical axes in van der Waals triclinic materials, Nat Commun **15**, (2024).

[42] G. Venturi, A. Mancini, N. Melchioni *et al.* Visible-frequency hyperbolic plasmon polaritons in a natural van der Waals crystal. Nat Commun **15**, 9727 (2024).

[43] Francesco L. Ruta et al. ,Good plasmons in a bad metal.Science 387,786-791(2025).

[44] H. Rubens and E. F. Nichols. Heat Rays Of Great Wave Length. Phys. Rev. (Series I) **4**, 314 (1897).

[45] L. D. LANDAU and E. M. LIFSHITZ, *CHAPTER I - ELECTROSTATICS OF CONDUCTORS*, in *Electrodynamics of Continuous Media (Second Edition)*, edited by L. D. LANDAU and E. M. LIFSHITZ, Vol. 8 (Pergamon, Amsterdam, 1984), pp. 1–33.

[46] G. R. Fowles, *Introduction to Modern Optics* (Holt McDougal, Evanston, IL, 1975).

[47] G. Álvarez-Pérez, K. V Voronin, V. S. Volkov, P. Alonso-González, and A. Y. Nikitin, Analytical approximations for the dispersion of electromagnetic modes in slabs of biaxial crystals, Phys Rev B **100**, 235408 (2019).

[48] See Supplementary Information.

[49] A. Y. U. NIKITIN, *Graphene Plasmonics*, in *World Scientific Handbook of Metamaterials and Plasmonics* (World Scientific, 2017), pp. 307–338.

[50] K. V. Voronin, G. Álvarez-Pérez, C. Lanza, P. Alonso-González, and A. Y. Nikitin, Fundamentals of Polaritons in Strongly Anisotropic Thin Crystal Layers, ACS Photonics **11**, 550 (2024).

[51] K. V. Voronin, G. Álvarez-Pérez, A. Tarazaga Martín-Luengo, P. Alonso-González and A. Y. Nikitin. Misalignment between the Directions of Propagation and Decay of Nanoscale-Confined Polaritons Nano Lett. 2025, 25, 37, 13811–13818

[52] Л. А. Фальковский, Оптические свойства графена и полупроводников типа $A_4B_6$, Uspekhi Fizicheskih Nauk **178**, 923 (2008).

[53] J. Chen, M. Badioli, P. Alonso-González, S. Thongrattanasiri, F. Huth, J. Osmond, M. Spasenovic, A. Centeno, A. Pesquera, P. Godignon, A. Zurutuza Elorza, N. Camara, F. J. García de Abajo, R. Hillenbrand & F. H. L. Koppens, Optical nano-imaging of gate-tunable graphene plasmons. Nature **487**, 77–81 (2012).

[54] C. Zheng, G. Hu, X. Liu, X. Kong, L. Wang, and C. W. Qiu, Molding Broadband Dispersion in Twisted Trilayer Hyperbolic Polaritonic Surfaces, ACS Nano **16**, 13241 (2022).

[55] C. Lanza, Twistoptics, GitLab repository, available at https://gitlab.com/ChristianLanza/twistoptics (2026).